\documentclass[pre,floats,longbibliography,twocolumn,superscriptaddress]{revtex4-1}

\usepackage{graphicx}
\usepackage{amsmath,amssymb,amsfonts,amsthm,enumerate}
\usepackage{rotate}
\usepackage{epsfig}
\usepackage[usenames,dvipsnames]{xcolor,colortbl}
\usepackage{multirow}
\usepackage{colortbl}
\usepackage[normalem]{ulem}
\usepackage{array}
\usepackage{arydshln}
\usepackage{booktabs}
\usepackage{url}
\usepackage{float}
\usepackage{paralist}
\usepackage{hyperref}

\newcommand{\etal}{\emph{et al.}}
\newcommand{\ie}{\emph{i.e.},}
\newcommand{\eg}{\emph{e.g.},}

\newcommand{\density}[1]{[ #1 ]}
\newcommand{\pathog}[1]{\sigma_#1}

\graphicspath{{pictures/}}

\begin{document}

\title{Emergence of synergistic and competitive pathogens in a co-evolutionary spreading model}

\author{Fakhteh Ghanbarnejad}
\email{fakhteh.ghanbarnejad@gmail.com}
\affiliation{Department of Physics, Sharif University of Technology, P.O. Box 11165-9161, Tehran, Iran}
\affiliation{Chair for Network Dynamics, Institute for Theoretical Physics and Center for Advancing Electronics Dresden (cfaed), Technical University of Dresden, 01062 Dresden, Germany}
\affiliation{Quantitative Life Sciences (QLS), The Abdus Salam International Centre for Theoretical Physics (ICTP), Strada Costiera, 11, I-34151 Trieste, Italy}

\author{Kai Seegers}
\affiliation{Institut f\"ur Theoretische Physik, Technische Universit\"at Berlin, Berlin, Germany}

\author{Alessio Cardillo}
\affiliation{Departament d'Enginyeria Inform\'atica i Matem\'atiques, Universitat Rovira i Virgili, Tarragona, Spain}
\affiliation{Laboratoire de Biophysique Statistique, \'Ecole Polytechnique F\'ed\'erale de Lausanne (EPFL), Lausanne, Switzerland}
\affiliation{GOTHAM Lab -- Instituto de Biocomputaci\'on y F\'isica de Sistemas Complejos (BIFI), Universidad de Zaragoza, Zaragoza, Spain}

\author{Philipp H\"ovel}
\affiliation{School of Mathematical Sciences -- University College Cork, Western Road, Cork, T12 XF62, Ireland}


\date{\today}


\begin{abstract}
Cooperation and competition between pathogens can alter the amount of individuals affected by a co-infection. Nonetheless, the evolution of the pathogens' behavior has been overlooked. Here, we consider a co-evolutionary model where the simultaneous spreading is described by a two-pathogen susceptible-infected-recovered model in an either synergistic or competitive manner. At the end of each epidemic season, the pathogens species reproduce according to their fitness that, in turn, depends on the payoff accumulated during the spreading season in a hawk-and-dove game. This co-evolutionary model displays a rich set of features. Specifically, the evolution of the pathogens' strategy induces abrupt transitions in the epidemic prevalence. Furthermore, we observe that the long-term dynamics results in a single, surviving pathogen species, and that the cooperative behavior of pathogens can emerge even under unfavorable conditions. 
\end{abstract}

\maketitle

\section{Introduction}
\label{sec:intro}

Understanding the diffusion of pathogenic agents is important as its aftermath reverberates on many aspects of our lives from health policies to economy, from politics to the transportation of people and goods \cite{anderson_book_1991,balcan_pnas_2009,eisenstein_nature_2016,bloom_pnas_2017,who_ihr_website}. Lately, the scientific community has devoted tremendous efforts in elucidating the dynamics behind these phenomena \cite{pastor_satorras_prl_2001,poletto_jtheobio_2013,pastor2015epidemic,heesterbeek_science_2015}. Despite the many achievements of computational epidemiology, the spreading of multiple pathogens has received less attention than expected. This comes as a surprise as phenomena like \emph{comorbidity} and \emph{cross-immunity} constitute the norm rather than the exception. The former indicates the simultaneous presence of multiple diseases within the same host. The latter denotes the acquisition of immunity towards a certain disease as a result of infection by another one. In the epidemiology jargon, pathogens supporting comorbidity are indicated as \emph{cooperators}, whereas the others as \emph{competitors}. Cooperative pathogens may mutually promote their contagion like the Spanish flu and pneumonia \cite{prasso2017postviral}, or HIV co-infections \cite{singer2017syndemics}. Some types of influenza are, instead, examples of competitors \cite{andreasen_jmatbio_1997,brundage2008deaths,oei2012relationship,sulkowski2008viral,sharma2005hiv,abu2006dual}. Moreover, comorbidity and cross-immunity are not observed exclusively among distinct pathogenic strains. For example, high levels of genetic diversity can provide a substrate for selection and rapid adaption, which are crucial to escape immune system recognition, developing resistance to drugs, and adapt to new host types (spillover) \cite{lloyd_smith_nature_2017}. Curiously, cases of cooperation have been reported also between distinct variants (quasi-species) of human H3N2 influenza \cite{xue_elife_2016,turner_nature_1999}. Consequently, the amount of literature about the spreading of cooperative \cite{sanz2014dynamics,cai2015avalanche,chen2013outbreaks,grassberger2016phase,chen2017phase,nucit_jstat_2017,masuda_jtb_2006} and competitive \cite{newman2005threshold,funk2010interacting,marceau2011modeling,miller2013cocirculation,poletto2015characterising,pinotti2020interplay} pathogens has grown over the years.\\
\indent Despite these evidences, the models developed hitherto assume that the \emph{strategy} of a pathogen to cooperate -- or not -- is \emph{costless}. In the evolutionary game framework, however, cooperation has a rather different meaning and implies always the payment of some \emph{costs} \cite{smith1982evolution,hofbauer1998evolutionary,nowak2006evolutionary}. For instance, the cooperative behavior of the Spanish flu exhibited towards secondary infections led to high death rates of the host subjects, making this synergistic epidemic -- or \textit{syndemic} -- prone to relatively ``quick'' disappearance \cite{prasso2017postviral}. Notably, the evolution of spreading pathogens has been rarely considered \cite{keeling2008modeling,roberts2002parasite,bohl-mol_biosys-2014,nurtay-chaos-2020}.\\
\indent In this paper, we envision an \emph{evolutionary} scenario where interacting pathogens have two different strains with cooperative and defective strategies, and evolve while maximizing their own benefit. We extend the two-strain susceptible-infected-recovered (SIR) model of Chen \etal{}~\cite{chen2013outbreaks} by intertwining it with an \emph{evolutionary infection game}, and show under which conditions synergy and competition emerge. We observe that the evolution of the pathogens' strategies reverberates on the epidemic prevalence. Moreover, the dynamics foster more the survival of single strains, \ie{} strategy, rather than the co-existence of multiple strains.\newline

\section{The model} 
\label{sec:model}

We combine two distinct processes: %
\begin{inparaenum}[(i)]
\item the simultaneous spreading of two different pathogens and
\item the evolution of their strategies.
\end{inparaenum}
The former takes place over a short, continuous, time-scale (within the season) indicated by variable $t$; whereas the latter occurs on a longer, discrete, one (between two seasons) denoted by variable $T$. The two processes are intertwined by using the outcome of the spreading as the input of the evolution and vice-versa (see Fig.~\ref{fig:model}). The two phases of spreading and evolution of strategies take place sequentially in a cyclic way until a global stationary state (GSS) is reached. We explain first the spreading and evolutionary game dynamics separately, and then describe how these processes (named phases $1$ and $2$) are combined together.

\subsection{Spreading dynamics} 
\label{ssec:spreading}

\begin{figure}[ht]
\centering
\includegraphics[width=0.98\columnwidth]{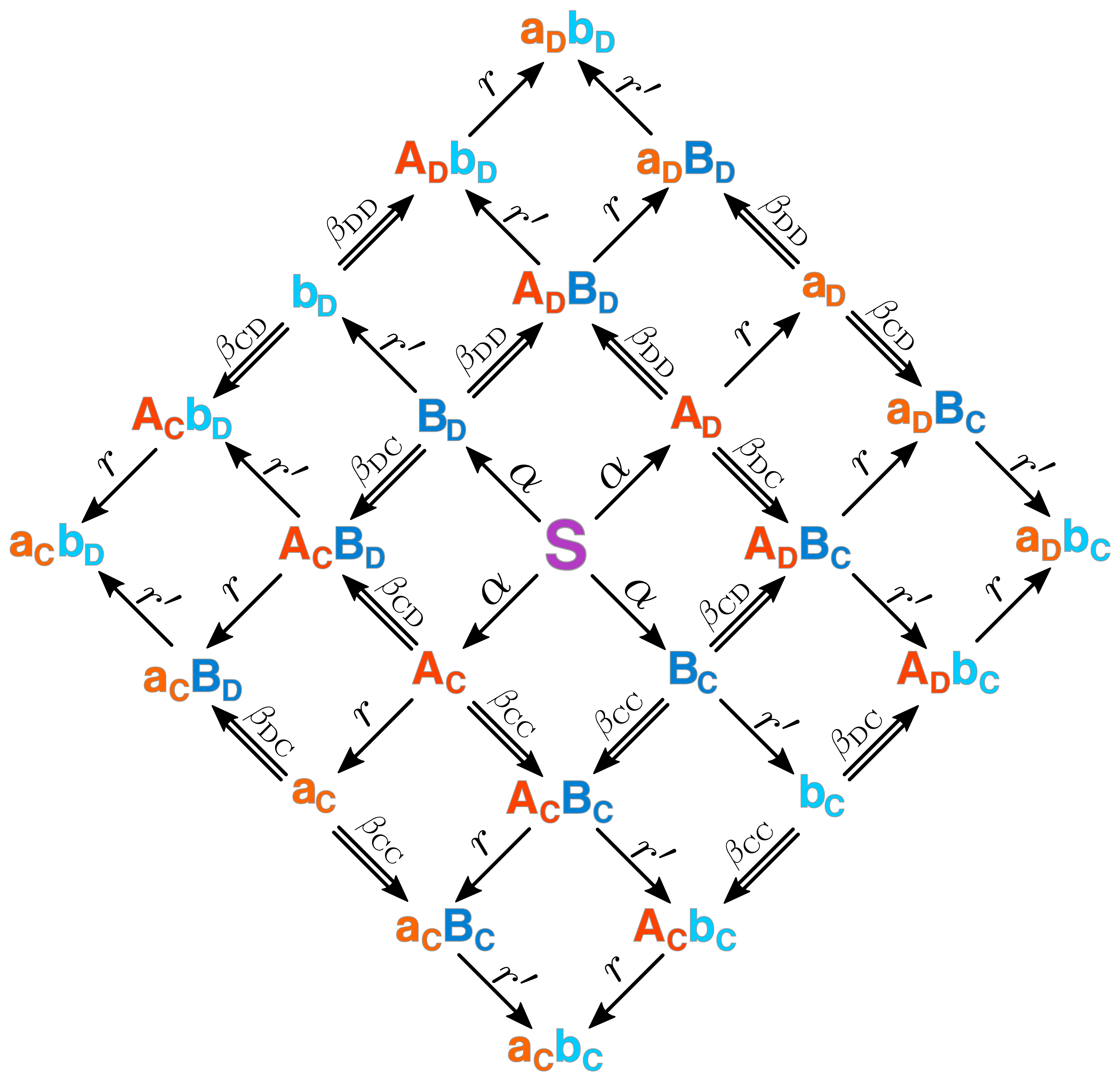}
\caption{Transition scheme of the two strategies double SIR spreading dynamics. We display all the possible transitions among compartments in the multi disease propagation of pathogens $A$ and $B$. The introduction of strategies for pathogens expands the original transition scheme introduced in \cite{chen2013outbreaks}, which now translates into a $25$ states diagram. Capital letters denote infected states, whereas small letters denote recovered ones. The infection transition is denoted by a single arrow for simple pathogen contagion, and by double arrows for multiple pathogens infection.}
\label{fig:transition_diag_game}
\end{figure}

In each season, the model describes the simultaneous spreading of two pathogens $A$ and $B$ that obey an extended SIR compartmental dynamics \cite{chen2013outbreaks}. Each pathogen, $\pathog{X}$, has two strains corresponding to its cooperator ($C$) and defector ($D$) strategies. 
\begin{description}
 \item[Cooperators ($\pathog{{X_C}}$)] A strain of pathogen $\pathog{X}$ that \emph{cooperates} with other strains.
 \item[Defectors ($\pathog{{X_D}}$)] A strain of pathogen $\pathog{X}$ that \emph{competes} with other strains.
\end{description}
Following the above definitions, the pathogenic population is composed by four different species, namely: ${A_C}$, ${A_D}$, ${B_C}$, and ${B_D}$. These species follow an SIR dynamics and interact with each other yielding a $25$ states model whose transitions' diagram is shown in Fig.~\ref{fig:transition_diag_game}.

%
%
%

In agreement with the case without strategies, and without loss of generality, the recovery rates are set equal to one (\ie{} $r = r^\prime = 1$). Moreover, we assume that the rates of infection depend only on the strategy of the pathogen occupying the host, which is equivalent to imposing a symmetry in the parameters' space. According to such a symmetry, we can write the infection rates of empty hosts as $\alpha_C = \alpha_D = \alpha$. Accordingly, for a host already occupied by one pathogen, we pass from four rates (\ie{} $\beta_{CC}$, $\beta_{CD}$, $\beta_{DC}$, and $\beta_{DD}$) denoting all the possible combinations of pairs of strategies to just two of them, namely: $\beta_{CC} = \beta_{DC} = \beta_C$ and $\beta_{DD} = \beta_{CD} = \beta_D$. After imposing the aforementioned symmetries, we can write the system of ODEs describing the spreading dynamics using an approach similar to that introduced in \cite{chen2013outbreaks}. Hence, we can define the groups of individuals actively infected by a species as:
\begin{equation}
 \label{eq:X}
 \begin{split}
    \!\!\!X_{A_C} &\!=\! \density{A_C} \!+\! \density{A_C B_C} \!+\! \density{A_C B_D} \!+\! \density{A_C b_C} \!+\! \density{A_C b_D} \\
    \!\!\!X_{A_D} &\!=\! \density{A_D} \!+\! \density{A_D B_C} \!+\! \density{A_D B_D} \!+\! \density{A_D b_C} \!+\! \density{A_D b_D} \\
    \!\!\!X_{B_C} &\!=\! \density{B_C} \!+\! \density{A_C B_C} \!+\! \density{a_C B_C} \!+\! \density{A_D B_C} \!+\! \density{a_D B_C} \\
    \!\!\!X_{B_D} &\!=\! \density{B_D} \!+\! \density{A_C B_D} \!+\! \density{a_C B_D} \!+\! \density{A_D B_D} \!+\! \density{a_D B_D} \,.
 \end{split}
\end{equation}
Capital letters denote infected states, whereas small letters denote recovered ones. The notation $\density{\cdot}$ indicates the density of individuals in a given state, while the sum of all $25$ states is normalized to $1$.

Considering the definitions of infected groups displayed in Eqs.~\eqref{eq:X}, we can write the complete set of equations modeling the spreading dynamics [Eqs.~\eqref{eq:25ode}]. Despite the high number of equations, the system in Eq.~\eqref{eq:25ode} depends only on three parameters: $\alpha$, $\beta_C$, and $\beta_D$. The next section illustrates how these parameters are related with each other.
%

%
%
\begin{figure*}[ht!]
\centering
\includegraphics[width=0.9\textwidth]{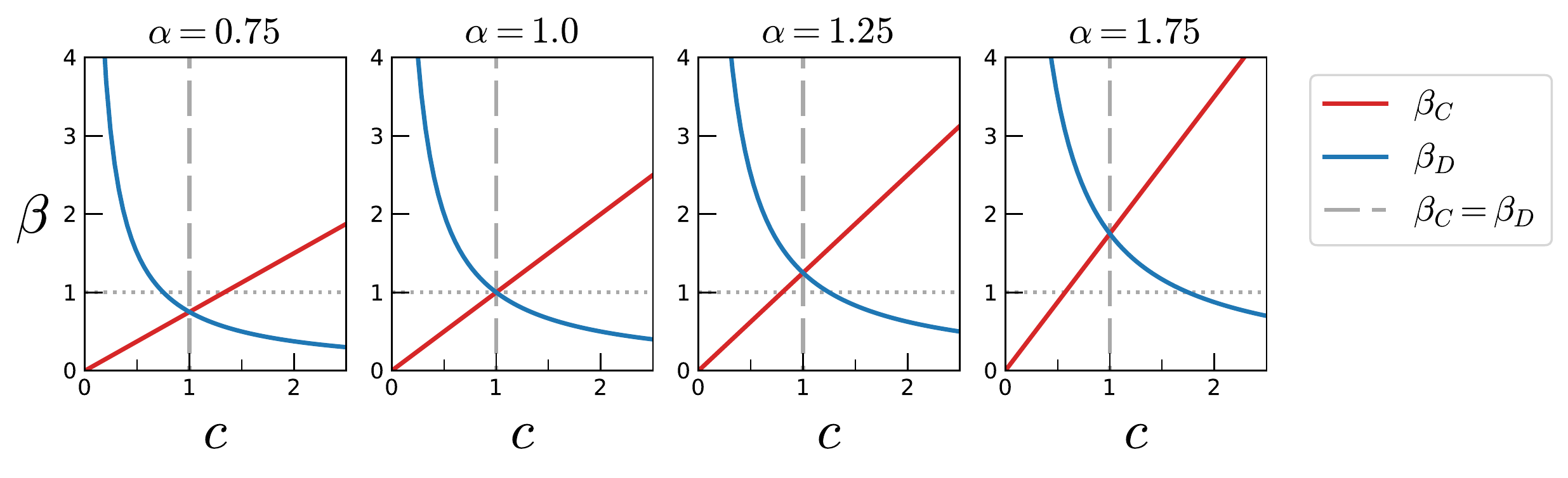}
\caption{Dependency of occupied host infection rate $\beta$ on the value of parameter $c$, according to Eq.~\eqref{eq:rates_relationships}, for four values of rate $\alpha$. The vertical dashed line denotes the value of $c$ at which $\beta_C = \beta_D = \alpha$. The horizontal dotted line at $\beta = 1$ separates the sub-critical region from the super-critical one.}
\label{fig:beta_vs_c}
\end{figure*}
%

%
%
{\allowdisplaybreaks
\begin{align}
\label{eq:25ode}
  \nonumber \dot{\density{S}} &= -\alpha \, \density{S} \, X_{A_C}  - \alpha \, \density{S} \, X_{A_D} - \alpha \, \density{S} \, X_{B_C} - \alpha \, \density{S} \, X_{B_D} \,, \\
  \nonumber \dot{\density{A_C}} &= \alpha \, \density{S} \, X_{A_C} - \beta_C \, \density{A_C} \, X_{B_C} - \beta_D \, \density{A_C} \, X_{B_D} - \density{A_C} \,, \\
  \nonumber \dot{\density{A_D}} &= \alpha \, \density{S} \, X_{A_D} - \beta_C \, \density{A_D} \, X_{B_C} - \beta_D \, \density{A_D} \, X_{B_D} - \density{A_D}  \,,\\
  \nonumber \dot{\density{B_C}} &= \alpha \, \density{S} \, X_{B_C}  - \beta_C \, \density{B_C} \, X_{A_C} - \beta_D \, \density{B_C} \, X_{A_D} - \density{B_C}  \,,\\
  \nonumber \dot{\density{B_D}} &= \alpha \, \density{S} \, X_{B_D} - \beta_C \, \density{B_D} \, X_{A_C} - \beta_D \, \density{B_D} \, X_{A_D} - \density{B_D} \,, \\
  \nonumber \dot{\density{A_C B_C}} &= \beta_C \, \density{A_C} \, X_{B_C} + \beta_C \, \density{B_C} \, X_{A_C} - 2 \, \density{A_C B_C} \,, \\
  \nonumber \dot{\density{A_C B_D}} &= \beta_D \, \density{A_C} \, X_{B_D} + \beta_C \, \density{B_D} \, X_{A_C} - 2 \, \density{A_C B_D} \,, \\
  \nonumber \dot{\density{A_D B_C}} &= \beta_C \, \density{A_D} \, X_{B_C} + \beta_D \, \density{B_C} \, X_{A_D} - 2 \, \density{A_D B_C} \,, \\
  \nonumber \dot{\density{A_D B_D}} &= \beta_D \, \density{A_D} \, X_{B_D} + \beta_D \, \density{B_D} \, X_{A_D} - 2 \, \density{A_D B_D} \,, \\
  \nonumber \dot{\density{a_C}} &= -\beta_C \, \density{a_C} \, X_{B_C} - \beta_D \, \density{a_C} \, X_{B_D} + \density{A_C} \,, \\
  \nonumber \dot{\density{a_D}} &= -\beta_D \, \density{a_D} \, X_{B_D} - \beta_C \, \density{a_D} \, X_{B_C} + \density{A_D} \,, \\
  \dot{\density{b_C}} &= -\beta_C \, \density{b_C} \, X_{A_C} - \beta_D \, \density{b_C} \, X_{A_D} + \density{B_C} \,, \\
  \nonumber \dot{\density{b_D}} &= -\beta_D \, \density{b_D} \, X_{A_D} - \beta_C \, \density{b_D} \, X_{A_C} + \density{B_D} \,, \\
  \nonumber \dot{\density{a_C B_C}} &= \beta_C \, \density{a_C} \, X_{B_C} + \density{A_C B_C} - \density{a_C B_C} \,, \\
  \nonumber \dot{\density{a_C B_D}} &= \beta_D \, \density{a_C} \, X_{B_D} + \density{A_C B_D} - \density{a_C B_D} \,, \\
  \nonumber \dot{\density{a_D B_C}} &= \beta_C \, \density{a_D} \, X_{B_C} + \density{A_D B_C} - \density{a_D B_C} \,, \\
  \nonumber \dot{\density{a_D B_D}} &= \beta_D \, \density{a_D} \, X_{B_D} + \density{A_D B_D} - \density{a_D B_D} \,, \\
  \nonumber \dot{\density{A_C b_C}} &= \beta_C \, \density{b_C} \, X_{A_C} + \density{A_C B_C} - \density{A_C b_C} \,, \\
  \nonumber \dot{\density{A_C b_D}} &= \beta_C \, \density{b_D} \, X_{A_C} + \density{A_C B_D} - \density{A_C b_D} \,, \\
  \nonumber \dot{\density{A_D b_C}} &= \beta_D \, \density{b_C} \, X_{A_D} + \density{A_D B_C} - \density{A_D b_C} \,, \\
  \nonumber \dot{\density{A_D b_D}} &= \beta_D \, \density{b_D} \, X_{A_D} + \density{A_D B_D} - \density{A_D b_D} \,, \\
  \nonumber \dot{\density{a_C b_C}} &= \density{A_C b_C} + \density{a_C B_C} \,, \\
  \nonumber \dot{\density{a_C b_D}} &= \density{A_C b_D} + \density{a_C B_D} \,, \\
  \nonumber \dot{\density{a_D b_C}} &= \density{A_D b_C} + \density{a_D B_C} \,, \\
  \nonumber \dot{\density{a_D b_D}} &= \density{A_D b_D} + \density{a_D B_D} \,.
\end{align}
} 

\subsection{Hierarchies between the rates of infection}
\label{ssec:alpha_and_betas_setup}

As mentioned above, the dynamics of the double SIR with cooperative and defective pathogens (\ie{} the quadruple SIR) depends on three parameters: the empty host infection rate $\alpha$, and the infection rates for hosts occupied either by a cooperator ($\beta_C$) or by a defector ($\beta_D$) pathogen. In the following, we show how these quantities are related to one another.

According to the definition of cooperative and defective strategies made in the previous section, it is natural to impose that $\beta_C > \beta_D$. Moreover, as discussed in \cite{cai2015avalanche}, the hierarchy between the empty host and occupied host infection rates determines whether two pathogens act in symbiosis (\ie{} their strategy is cooperation) or in antagonism (\ie{} their strategy is defection). The former case occurs when $\alpha < \beta$, whereas the latter when $\alpha > \beta$. These hierarchies between the infection rates denote the existence of a relationship between them. Amidst the plethora of possible choices, we decided to adopt the following one:
\begin{equation}
\label{eq:rates_relationships}
\begin{split}
\beta_C &= \alpha \, c \,,\\
\beta_D &= \frac{\alpha}{c}\,,
\end{split}
\end{equation}
with $c \in \, ] 0, \infty [$ being a parameter. Imposing a relationship between the infection rates reduces the number of free parameters in Eqs.~\eqref{eq:25ode} from three to two: $\alpha$ and $c$.

\begin{figure}[ht!]
\centering
\includegraphics[width=0.8\columnwidth]{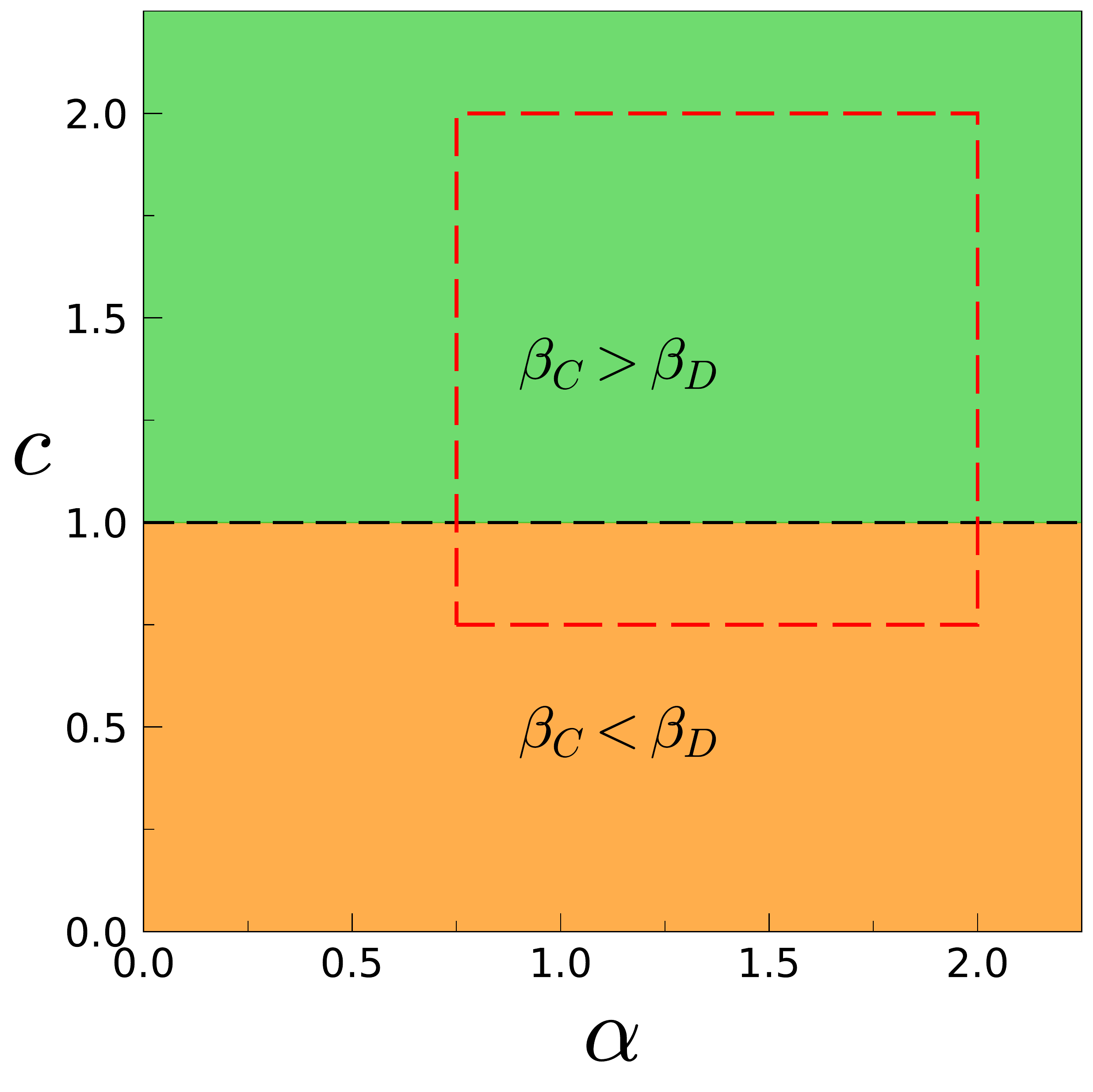}
\caption{Partition of the $(\alpha,c)$ space according to the hierarchy between $\beta_C$ and $\beta_D$. The area delimited by the red dashed square corresponds to the area portrayed in panels (a) and (c) of Fig.~\ref{fig:dynamics_alphaplot_cplot}.}
\label{fig:beta_hierarchy_alpha_c_space}
\end{figure}

Figure \ref{fig:beta_vs_c} displays the values of $\beta_C$ and $\beta_D$ as a function of $c$ for four different values of $\alpha$: two sub-critical (\ie{} $\alpha \leq 1$) and two super-critical (\ie{} $\alpha > 1$). The visual inspection of Fig.~\ref{fig:beta_vs_c} reveals that, in agreement with Eq.~\eqref{eq:rates_relationships}, for $c=1$ one gets $\beta_C = \beta_D = \alpha$. Such a case corresponds to the \emph{neutral spreading scenario} in which the rate of infection is independent on the occupation state of the host, as well as on the strategy of the pathogen occupying it. Hence, the only factor influencing the outcome of the dynamics is the value of $\alpha$ which controls the epidemic prevalence $1 - S_\infty$. The value $c=1$ corresponds also to the point where the hierarchy between $\beta_C$ and $\beta_D$ changes. In Fig.~\ref{fig:beta_hierarchy_alpha_c_space} we plot the hierarchy between $\beta_C$ and $\beta_D$ as a function of $\alpha$ and $c$. In particular, the region $c < 1$ corresponds to the case $\beta_D > \beta_C$ corresponding to the scenario in which it is easier to infect a host occupied by a pathogen acting as a defector than a cooperator. Although such a scenario is biologically not meaningful, we nevertheless explore it for the sake of completeness.

%
%

In summary, assuming that it is easier to infect a host already infected by a cooperator pathogen than by a defector one, we set $\beta_{CC} = \beta_{DC} = \beta_C = \alpha \, c$ and $\beta_{DD} = \beta_{CD} = \beta_D = \tfrac{\alpha}{c}$ with $c > 0$.

\subsection{Evolutionary game dynamics}
\label{ssec:evol_games}

The quadruple SIR model encoded by Eqs.~\eqref{eq:25ode} and the hierarchies between the infection rates analyzed in Sec.~\ref{ssec:alpha_and_betas_setup} describe the spreading of pathogens acting in symbiosis or antagonism with the other pathogens. However, the sole consequence of adopting a certain strategy is how easily a pathogen can infect a host occupied by another pathogen. Nevertheless, from the biological point of view, facilitating (or not) the infection from another pathogen could be thought -- among other things, -- as a proxy for the will of the occupying pathogen to share (or not) the host's resources with the invader.

Following this hypothesis, one could model the act of infecting a host as a game and, consequently, assign a payoff to it \cite{hofbauer1998evolutionary}. The payoff accumulated by all the pathogens of species $X$ and strategy $Y$, $\Pi_{\pathog{{X_Y}}}$, constitutes the \emph{fitness} of that pathogen's type which, in turn, determines its ability to reproduce. Hence, we can use the fitness of each pathogen's type computed at the end of the $T$-th season of the spreading process to compute its abundance in the initial seed of the $(T+1)$-th season. Such an approach allows to describe features like comorbidity and cross-immunity as the byproduct of natural selection.

In light of the above reasoning, the act of infecting a host splits into two main scenarios: one in which the host is empty, and another in which it is already occupied by another pathogen. Assuming that the total amount of resources available in the host is equal to one, let us discuss the two scenarios separately.

If the host is empty, then the pathogen infecting the host will have access to all of its resources regardless of its strategy. Therefore, the payoff, $\pi$, associated to the event of infecting an empty host (\ie{} single infection) is equal to $\pi = 1$. If the host is occupied instead (\ie{} in the secondary infection event), the value of the payoff depends on the combination of the strategies of the pathogen infecting the host, and of the pathogen already present within the host. Since we assume that pathogens can either cooperate or defect, and that cooperators are more keen to share the host's resources than defectors, we can adopt the payoff matrices of the Hawk and Dove (HD) game \cite{smith1982evolution,szabo_physrep_2007}, given by:
\begin{equation}
\label{eq:payoff_matrix_hd}
\mathcal{A} = 
\bordermatrix{~ & C & D \cr
                  C & \frac{1}{2} & \gamma \cr
                  D & 1 - \gamma & -\frac{1}{2}} %
\qquad
\mathcal{A}^\prime = 
\bordermatrix{~ & C & D \cr
                  C & \frac{1}{2} & 1 - \gamma \cr
                  D & \gamma & -\frac{1}{2}} \, ,
\end{equation}
with $\gamma \in \left[ 0, \tfrac{1}{2} \right]$. The payoff, $\pi_{X,Y}$, of a pathogen with strategy $X$ infecting a host infected by another pathogen with strategy $Y$ corresponds to the element $a_{X,Y}$ ($X,Y \in \{C, D\}$) of $\mathcal{A}$. Analogously, the payoff of a pathogen with strategy $Y$ occupying a host that gets infected by another pathogen with strategy $X$ corresponds to the element $a^\prime_{X,Y}$ of $\mathcal{A}^\prime$. The fitness of species $\sigma$, $\Pi_\sigma$, depends on the history (\ie{} the sequence) of its contagion record (events). Finally, the concentration, $\rho_i$, of species $i$ in the spreading seed of season $T+1$ is regulated by the so-called \emph{replicator equation} \cite{hofbauer1998evolutionary, nowak2006evolutionary}, given by:
\begin{equation}
\label{eq:replicator}
\left. \rho_i^{T+1} \right\vert_{t_0} = \Bigl. \rho_i^T \Bigr\vert_{t_0} \biggl[ 1 + \Bigl. \Pi_i^T \Bigr\vert_{t_\infty} \!\!- \Bigl. \overline{\Pi}^T \Bigr\vert_{t_\infty} \biggr]\qquad 
\end{equation}
where $i = \{ A_C, A_D, B_C, B_D \}$ and $\bigl. \rho_i^{T} \bigr\vert_{t_0}$ is the concentration of species $i$ at the beginning ($t_0$) of spreading season $T$. $\bigl. \Pi_i^T \bigr\vert_{t_\infty}$ is the \emph{fitness} of species $i$ equivalent to the total payoff accumulated during spreading season $T$. Finally, $\bigl. \overline{\Pi}^T \bigr\vert_{t_\infty}\!\!$ is the average of the fitness obtained during season $T$ by all species $\bigl. \overline{\Pi}^T \bigr\vert_{t_\infty} \!\!\!\!= \dfrac{1}{4} \sum_i \bigl. \Pi_i^T \bigr\vert_{t_\infty}\!\!$, and $t_\infty$ denotes the time $t$ at which the spreading dynamic reached its stationary state. According to Eq.~\eqref{eq:replicator}, species with fitness higher than the average will proliferate, whereas those with fitness lower than the average will become extinct \cite{nowak2006evolutionary}. 

%
%
%
%
\begin{figure*}[ht]
\centering
\includegraphics[width=0.92\textwidth]{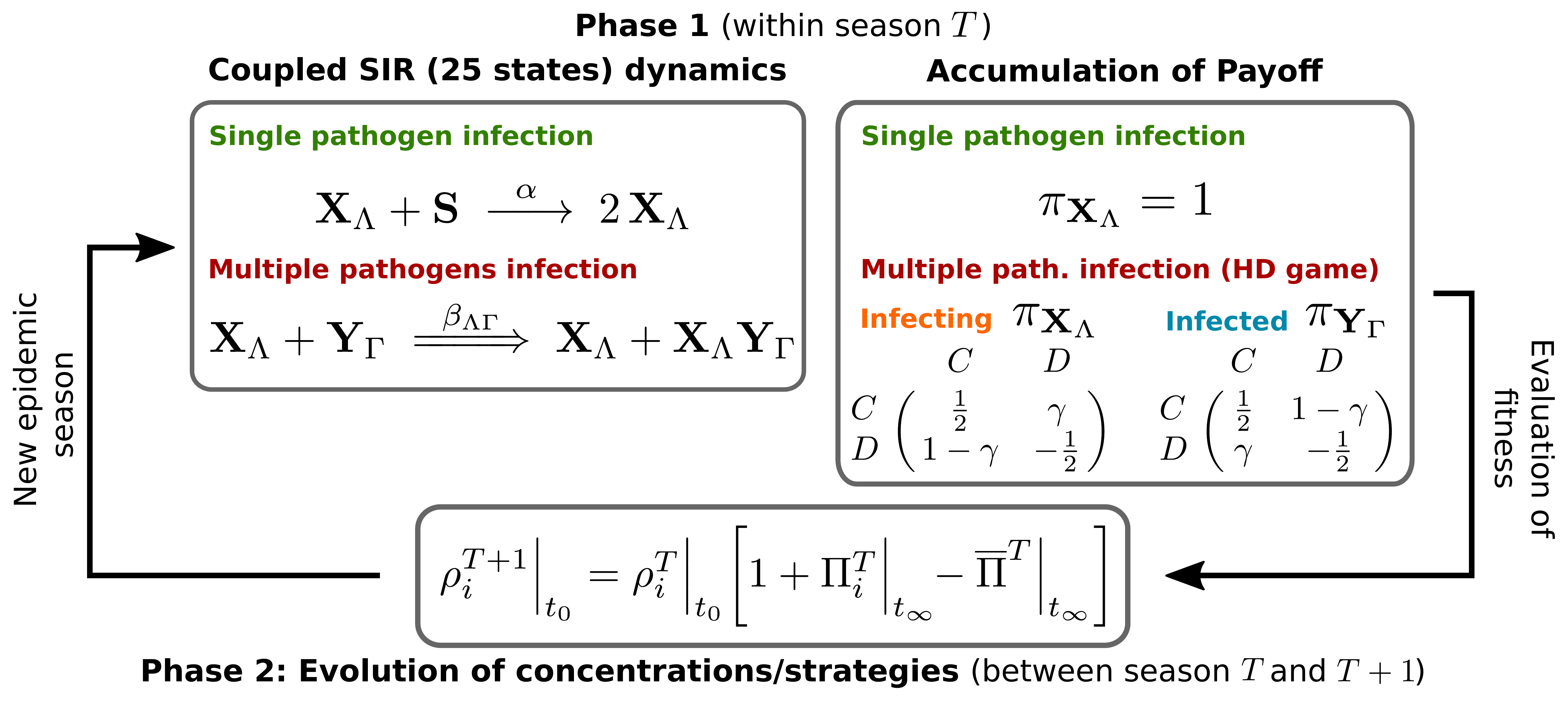}
\caption{Schematic representation of the co-evolutionary model. Phase 1 describes the processes occurring during each spreading season, $T$. The left box depicts the single and multiple SIR reaction kinetics. The right box shows the corresponding accumulation of payoff, $\pi$, by species $\mathbf{X}_\Lambda$ and $\mathbf{Y}_\Gamma$ with $\mathbf{X}, \mathbf{Y} \in \{A, B\}$, and strategy $\Lambda, \Gamma \in \{ C, D\}$. After spreading season $T$ ends, the concentrations, $\rho$, of species in the initial seed evolves following the replicator equation [Eq.~\eqref{eq:replicator}] (phase 2). The two phases repeat several times until the global stationary state (GSS) is reached. The parameters do not change during the dynamics.}
\label{fig:model}
\end{figure*}

\subsection{The co-evolutionary model}
\label{ssec:coevol_model}

Figure~\ref{fig:model} summarizes the main features of our co-evolutionary dynamics model, in which the spreading and the evolutionary game dynamics introduced previously correspond to the so-called phases $1$ and $2$ of our model.

At the start of each season (\ie{} $t = t_0 = 0$), the epidemic seed is composed exclusively of a mixture of species $A_C$, $A_D$, $B_C$, and $B_D$. The initial pathogen concentrations, $\rho_i$, ($i \in \{ A_C, A_D, B_C, B_D \}$) evolve according to Eq.~\eqref{eq:replicator} (phase 2). Without losing generality, we fix the size of the total initial infectious seed for all seasons \ie{} $\sum_i \, \bigl. \rho_i^T\bigr\vert_{t_0} \!\! = 0.05 \; \forall \, T$. During each spreading season, $T$, the different pathogen species spread according to Eqs.~\eqref{eq:25ode} (whose reaction kinetics is summarized in the top left panel of Fig.~\ref{fig:model} and details are shown in Fig. \ref{fig:transition_diag_game}) and accumulate payoff via infection events according to the infection game introduced in Sec.~\ref{ssec:evol_games} (Fig.~\ref{fig:model} top right panel). Specifically, a pathogen $\mathbf{X}_\Lambda$ infecting a susceptible host $\mathbf{S}$ receives a payoff $\pi_{\mathbf{X}_\Lambda} = 1$ regardless of its strategy. If, instead, the host is already infected by another pathogen $\mathbf{Y}_\Gamma$ -- \ie{} in a secondary infection, -- the payoff of the infecting pathogen, $\pi_{\mathbf{X}_\Lambda}$, is given by the $(\Lambda, \Gamma)$ element of matrix $\mathcal{A}$ of Eq.~\eqref{eq:payoff_matrix_hd} according to the four possible combinations of the (infecting, infected) pathogens' strategies: $(C,C)$, $(C,D)$, $(D,C)$, and $(D,D)$ (similar reasoning applies to payoff $\pi_{\mathbf{Y}_\Gamma}$ using matrix $\mathcal{A}^\prime$). The choice of the payoff scheme (\ie{} the game) associated with the secondary infection event is dictated by the idea of how the pathogens share the host's resources. Moreover, the idea that cooperators will share the host's resources while defectors will try to fight to seize them all, reverberates also on the properties of the secondary infection rates $\beta_C$ and $\beta_D$ which: %
\begin{inparaenum}[i)]
\item depend exclusively on the strategy of the pathogen already occupying the host, and
\item controls the hierarchy existing among them (\ie{} $\beta_C > \beta_D$).
\end{inparaenum}
For these reasons, a payoff scheme like that of the HD game fits quite well with our model. Considering other payoff schemes like the Stag-Hunt one which are typical of ``\emph{coordination dynamics}'' (\ie{} where the best strategy is to coordinate with the other player and adopt the same strategy) clashes with the idea of having finite host's resources. A similar reasoning applies to the case of the Prisoner's Dilemma (PD) \cite{nowak2006evolutionary}.

Once the spreading process reaches its stationary state, \ie{} $t = t_\infty = \infty$, the values of $\rho$, for the next season $T+1$ are given by Eq.~\eqref{eq:replicator} (Fig.~\ref{fig:model}, phase 2). The mixture of species in the seed evolves season after season until $\rho_i^T$ reaches a stationary state, \ie{} $\rho_i^{T+1} = \rho_i^{T} \; \forall i$. Then, the co-evolutionary dynamics is considered at the GSS which is labeled as evolutionary season $T = \infty$.

\section{Results} 
\label{sec:results}

\subsection{Differences between non evolutionary and evolutionary dynamics}
\label{ssec:res_host_pathogen_perspective}

\begin{figure*}[ht!]
\centering
\includegraphics[width=0.9\textwidth]{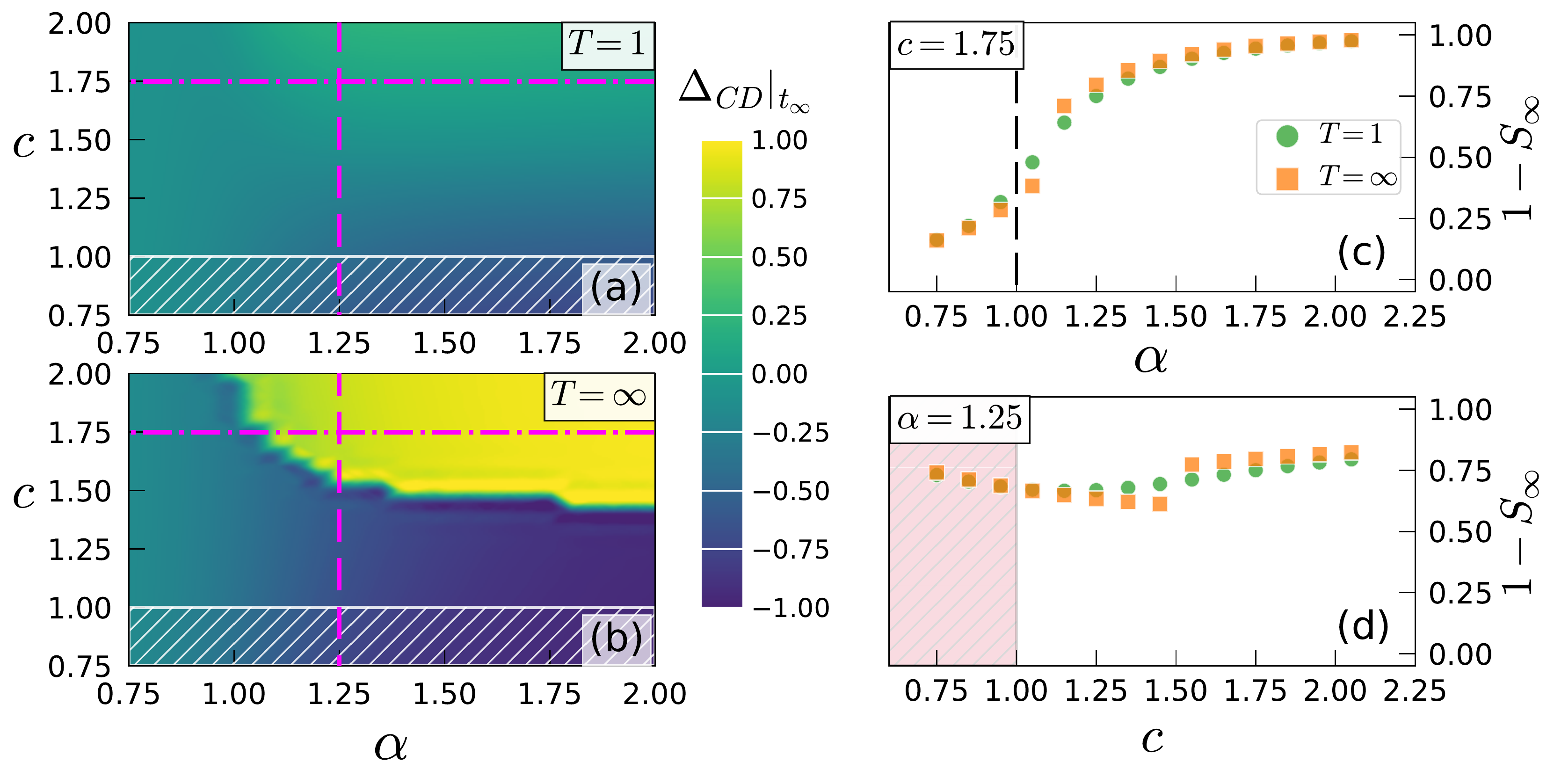}
\caption{Characterization of the co-evolutionary dynamics. Pathogens' perspective: Panels (a) and (b) display the value of strategy's balance $\Delta_{CD}\vert_{t_\infty}$ without evolution [$T=1$, panel (a)], and at the GSS [$T=\infty$, panel (b)] as a function of $\alpha$ and $c$. Brighter colors denote populations with a higher prevalence of cooperators. The slanted area denotes the region of the $(\alpha,c)$ space with $\beta_D > \beta_C$. Host's perspective: Panels (c) and (d) report the values of disease incidence $1 - S_\infty$ without evolution ($T=1$) and at the GSS ($T=\infty$) as a function of $\alpha$ and $c$, respectively. The other parameter's value is fixed as indicated by either the horizontal or vertical line in panels (a) and (b). The shaded area in panel (d) corresponds to the regime $\beta_D > \beta_C$ (\ie{} $c<1$). Other parameters: $\gamma = 0.25$, $\sum_i \rho_i^T|_{t_0}=0.05$, and $\Delta _{CD}|_{t_0}=-0.03$.}
\label{fig:dynamics_alphaplot_cplot}
\end{figure*}

We start our analysis by looking at the state of the system at the end of the spreading dynamics, $t_{\infty}$, and compare it at the first ($T=1$) and the last ($T=\infty$) season. In Fig.~\ref{fig:dynamics_alphaplot_cplot} we portray the outcome of the co-evolutionary dynamics from both the pathogens (left panels) and the host (right panels) perspectives. If the composition of the spreading seed, $\Delta_{CD}\vert_{t_0} = \bigl( [A_C] + [B_C] \bigr) - \bigl( [A_D] + [B_D] \bigr)$, is richer in cooperation, \ie{} $\Delta_{CD}\vert_{t_0} > 0$, then cooperator strains will easily take over.
%
%
%
%

Nevertheless, there is a region of the $(\alpha, c)$ space where cooperation still has the chance to thrive even though the seed is dominated by defectors (\ie{} $\Delta_{CD}\vert_{t_0} < 0$). Such regions correspond to the brighter areas displayed in Figs.~\ref{fig:dynamics_alphaplot_cplot}(a) and (b), where the colors encode the density of the pure cooperator and defector strains at the end of the spreading season, $\Delta_{CD}\vert_{t_\infty} = \Bigl( [a_C] + [b_C] + [a_C b_C] \Bigr) - \Bigl( [a_D] + [b_D] + [a_D b_D ] \Bigr)$. Note that the lower case variables refer to the recovered compartments. We find that the co-evolutionary dynamics amplifies the differences of $\Delta_{CD}\vert_{t_\infty}$ observed for the first season $T=1$ [panel (a)], and a region where cooperation thrives emerges, \ie{} the bright area in panel (b) corresponding to $\Delta_{CD}\vert_{t_\infty} \!\! > 0$. In addition, we observe another region where defection prevails ($\Delta_{CD}\vert_{t_\infty} \!\! < 0$).

Figures~\ref{fig:dynamics_alphaplot_cplot}(c) and (d) capture the effects of the evolution from the hosts' perspective through the lens of disease incidence (or transitivity), $1 - S_\infty$. As we vary $\alpha$ [panel (c)], we observe a transition from lower incidences for $\alpha \leq 1$, towards complete infection, \ie{} $1 - S_\infty \simeq 1$. In the absence of evolution, that is, for a single season only ($T=1$), the transition between these regimes is smooth, whereas the catalytic effect of the evolutionary dynamics, \ie{} over many seasons ($T = \infty$), triggers the appearance of a gap in the transition. Even though at first glance the gap resembles the discontinuous transitions observed in the cooperative SIR co-infection models \cite{chen2013outbreaks, zarei2019exact}, they are not the same. In fact, in non-evolutionary SIR co-infection models the gap occurs exclusively for cooperator only pathogens, whereas purely defector pathogens do not make any gap (see the competitive regime in Ref.~\cite{chen2013outbreaks}). Instead, in our case both cooperators and defectors spread at the same time, thus highlighting the inability of the latter to suppress the gap. Also, the transition occurs at smaller values of $c$ and greater values of $\alpha$, compared to cooperative only SIR dynamics, (see Eq.~(7) in Ref.~\cite{zarei2019exact}).

The effect of varying $c$ on the disease incidence [panel (d)] is less strong. Notwithstanding, we still observe a gap in the transition for $T = \infty$. In the neutral spreading setup $\beta_C = \beta_D = \alpha$ (\ie{} $c=1$), we do not observe any difference between the incidences measured in the absence and presence of seed evolution. Such a similarity suggests that the sole accumulation of payoff is not enough to induce differences in the phenomenology. Finally, we stress that the region corresponding to $c < 1$ is biologically not meaningful as $\beta_C < \beta_D$.

\subsection{Evolution of the strategies across seasons}
\label{ssec:res_evolution_of_strategies}

%
%
%
\begin{figure}[ht!]
\centering
\includegraphics[width=0.8\columnwidth]{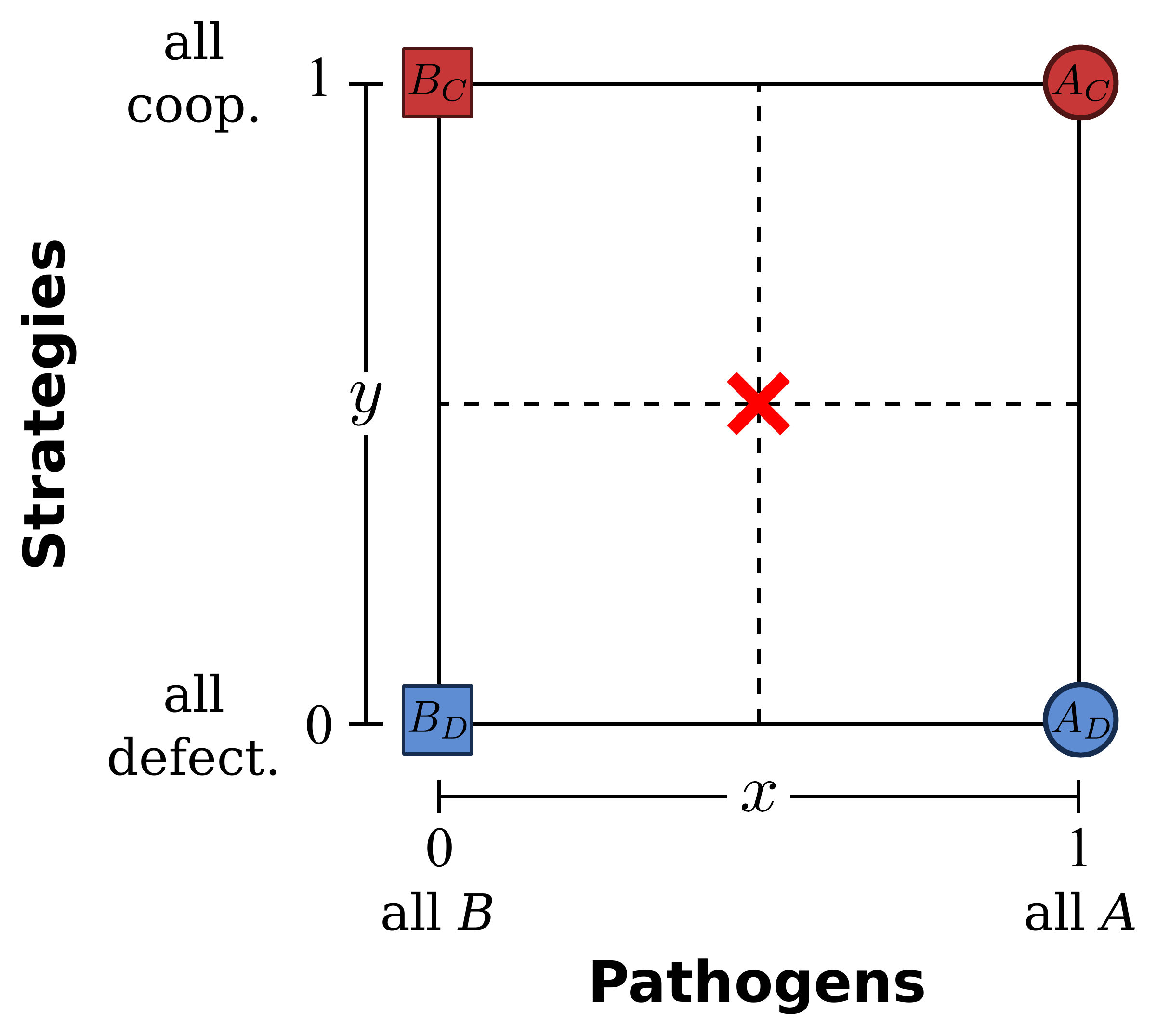}
\caption{Schema of the bidimensional projection of the four-dimensional hypercube describing the composition of the spreading seed in terms of the densities of each pathogen. The $x$ coordinate accounts for the pathogen's specie, whereas the $y$ coordinate accounts for the pathogen's strategy.}
\label{fig:bidimensional_projection_scheme}
\end{figure}

A way to study the evolution of the dynamics is to look at the evolution across the seasons of the spreading seed. More specifically, the sum of the densities of pathogens' species at the beginning of each spreading season -- \ie{} the size of the spreading seed, -- is fixed and equal to $\omega \in \, ]0, 1[$. As the pathogens' densities are all positive definite, and their sum is constant, the equation:
\begin{equation}
\label{eq:seed_conservation}
\bigl.\rho_{A_C}\bigr\vert_{t=0} + \bigl.\rho_{A_D}\bigr\vert_{t=0} + \bigl.\rho_{B_C}\bigr\vert_{t=0} + \bigl.\rho_{B_D}\bigr\vert_{t=0} = \omega \,,
\end{equation}
describes a three-dimensional hyperplane on a four-dimensional hypercube (\ie{} a tesseract) corresponding to the $\left\lbrace \rho_{A_C}, \rho_{B_C}, \rho_{A_D}, \rho_{B_D} \right\rbrace$ space. Such an object and trajectories on it are not always straightforward to visualize on a bidimensional surface. Nevertheless, under certain conditions it is possible to project the density space onto a bidimensional surface. To this aim, let us consider two variables $0 \leq x \leq 1$ and $0 \leq y \leq 1$ fulfilling the following conditions:
\begin{equation}
\label{eq:four_dimension_coordinates}
\begin{split}
\Bigl.\rho_{A_C} = \density{A_C}\Bigr\vert_{t=0}  &= \omega \, x \, y \,, \\
\Bigl.\rho_{A_D} = \density{A_D}\Bigr\vert_{t=0}  &= \omega \, x  \left( 1 - y \right) \,,\\
\Bigl.\rho_{B_C} = \density{B_C}\Bigr\vert_{t=0}  &= \omega  \left( 1 - x \right) \, y \,,\\
\Bigl.\rho_{B_D} = \density{B_D}\Bigr\vert_{t=0}  &= \omega  \left( 1 - x \right) \left( 1 - y \right) \,.
\end{split}
\end{equation}

The $x$ coordinate accounts for the pathogen's species concentration with $x=0$ ($x=1$) denoting a population fully made of pathogens of species $B$ ($A$). The $y$ coordinate, instead, accounts for the pathogen's strategy, with $y=0$ ($y=1$) denoting full defection (cooperation). Not all the possible combinations of the seed components (only three of them are truly independent) can be mapped using the present approach. Still, the bidimensional representation allows one to scrutinize the dynamics between the most relevant points. Figure~\ref{fig:bidimensional_projection_scheme} provides a visual summary of the bidimensional projection.

The diagram's corners correspond to pure seeds (\ie{} made of only one species with one strategy), with $\density{A_C} = \omega$ corresponding to the point with coordinates $(1, 1)$, $\density{A_D} = \omega$ to $(1, 0)$, $\density{B_C} = \omega$ to $(0, 1)$, and $\density{B_D} = \omega$ to $(0, 0)$, respectively (we are omitting the $t=0$ subscript). The middle points on the edges of the $[0,1] \times [0,1]$ square indicate, instead, a seed balanced in either the pathogen's species (\eg{} $\density{A_C} = \density{A_D} = \tfrac{\omega}{2}$ with coordinates $(\tfrac{1}{2}, 1)$) or strategy (\eg{} $\density{A_C} = \density{B_C} = \tfrac{\omega}{2}$ with coordinates $(0, \tfrac{1}{2})$). Finally, the red cross at the middle of the diagram corresponds to the case $\density{A_C} = \density{B_C} = \density{A_D} = \density{B_D} = \tfrac{\omega}{4}$ with coordinates $(\tfrac{1}{2}, \tfrac{1}{2})$. Therefore, the path connecting point $P \equiv (x_P, y_P)$ with point $Q \equiv (x_Q, y_Q)$ denotes a variation in the composition of the spreading seed induced by the evolutionary component of our model.

\begin{figure*}[ht!]
\centering
\includegraphics[width=0.98\textwidth]{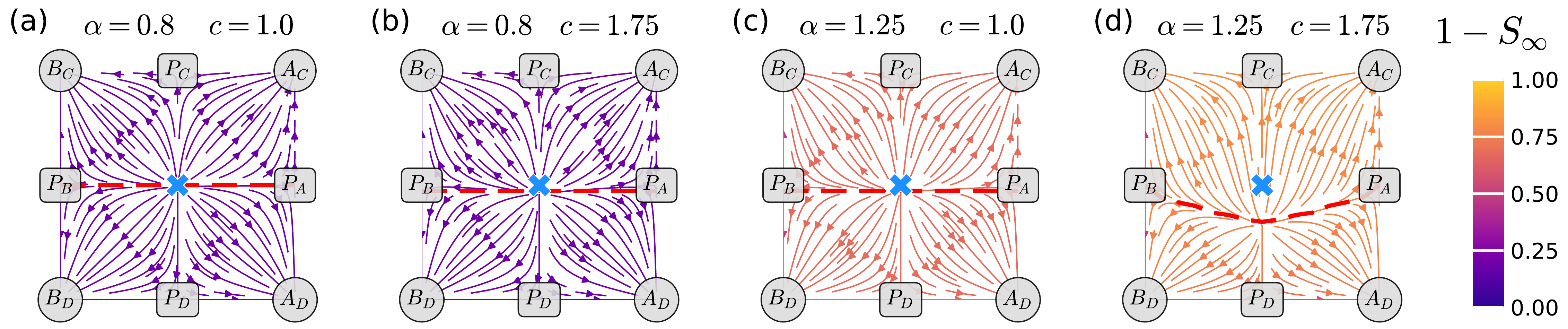} 
\caption{Evolution of the epidemic seed as two-dimensional projections of stream plots. Pure states correspond to the corners of the diagram, while strategy or species balanced states ($P_X$) are located at the middle of the borders, (\eg{} $P_C \equiv [A_C] = [B_C]$). The central point (blue cross) denotes the mixed initial state where $[A_C]_{t_0} = [A_D]_{t_0} = [B_C]_{t_0} = [B_D]_{t_0} = \tfrac{1}{4}\sum_i \rho_i^T|_{t_0}$. The colors encode the transitivity at the GSS. The red, dashed line is an approximation to guide the eye finding the manifold separating cooperation from defection. The arrows indicate the displacement of the seed between epidemic seasons. We set $\sum_i \rho_i^T|_{t_0} = 0.05$ and $\gamma=0.25$, and use four sets of $(\alpha, c)$ pairs. See Fig. \ref{fig:bidimensional_projection_scheme} of  for the details of the bi-dimensional mapping.} 
\label{fig:phase_portrait_GSS}
\end{figure*}

We continue our analysis by studying the evolution of the epidemic seeds between seasons, to investigate how the interplay between the two phases of our model impacts the overall dynamics. To this aim, Fig.~\ref{fig:phase_portrait_GSS} displays how the concentrations of species $\rho_{A_C}, \rho_{B_C}, \rho_{A_D}$, and $\rho_{B_D}$ in the seed change season after season until reaching the GSS.

%
%
%
%

The main observations are summarized as follows:
\begin{itemize}

\item Contrary to the expected behavior of the HD game, the pure states correspond to stable fixed points of the dynamics, whereas the balanced states are saddle points, \ie{} they are stable in one direction and unstable in the other. Hence, no matter what the initial composition of the seed is, only one species with one strategy survives at the GSS, while all other species become extinct.
\item Two unstable manifolds split the phase portrait according to each of the pathogens' features, \ie{} species and strategy.
\item Figures~\ref{fig:phase_portrait_GSS}(a) and (b) show that, when the simple infection rate is sub-critical (\ie{} $\alpha < 1$), the structure of the phase portrait and the epidemic prevalence are not affected by variations of $c$.
\item For supercritical values of the simple infection rate [$\alpha > 1$, \ie{} Figs.~\ref{fig:phase_portrait_GSS}(c) and (d)], increasing $c$ translates into an expansion of cooperation's basin of attraction. Hence, seeds initially composed of more defectors than cooperators can evolve towards fully cooperative strains.
\item Even though changing the value of $\gamma$ displaces the intermediate stable fixed point in the ``classical'' HD game dynamics; we have verified that both the position of the unstable manifolds and the epidemic prevalence are robust to variations of $\gamma$.
\end{itemize}

\section{Discussion and conclusion} In this paper, we have studied the emergence of synergistic and competitive traits in a co-evolutionary model intertwining epidemic spreading and evolutionary game theory. To this aim, we have extended the SIR model of Chen \etal{} \cite{chen2013outbreaks} by assuming that pathogens have two strategies (cooperate or defect), and allowing that their concentration inside the epidemic seed (whose total size is kept fixed throughout the whole dynamics) evolves according to the replicator equation \cite{smith1982evolution}. For the latter, the pathogens accumulate payoff according to an infection game whose payoff scheme for secondary infections events maps onto the hawk-and-dove game.\newline
\indent The resulting co-evolutionary model displays features that are not present in either spreading or evolutionary game dynamics taken singularly. In particular, the game reverberates on the outcome of the spreading dynamics. It alters the disease prevalence in the host and can trigger the emergence of a gap in the transition. This phenomenon has also been observed for non-evolutionary SIR models where two pathogens only cooperate and sharp transitions occur for $c > 2$ and $\alpha < 1$ \cite{chen2013outbreaks,zarei2019exact}. Moreover, the interplay between the spreading and game dynamics alters the stability of the state space's fixed points of the game. Pure seeds are stable solutions of the dynamics, contrary to the expected stable \emph{mixed} scenario typical of the hawk-and-dove game~\cite{nowak2006evolutionary}. We have also observed the emergence of a region of the phase space where cooperation thrives even under unfavorable conditions. Such a behavior is in agreement with the phenomenology reported for stochastic games where cooperation emerges even in conditions where usually it should not \cite{stollmeier-prl-2018,hilbe-nature-2018,szolnoki-scirep-2019}. 

Summing up, the overall phenomenology exhibited by our model cannot be traced back to one of the two processes separately but emerges from their interplay.
The framework presented here constitutes a way to understand under which conditions synergy, \ie{} comorbidity, and competition, \ie{} cross-immunity, can emerge through an evolutionary framework where the hosts and the pathogens mutually interact. The present work is a \emph{first attempt} towards understanding multiple pathogen spreading and opens new roads for theoretical modeling of multistrain epidemic processes from the perspective of evolutionary theory. Despite the richness of the dynamical scenarios observed, we made simplifying assumptions. For instance, we have considered a symmetric scenario for the primary infection rates and have chosen a specific set up for the secondary infection rates. Moreover, the values of the parameters do not evolve across time, and we have considered a well mixed population. Relaxing these assumptions can lead to even more realistic dynamics.

\section*{acknowledgments}

The authors thank the members of the GOTHAM lab (and in particular D.~Soriano Pa\~nos) for helpful comments and discussions. AC acknowledges the financial support of SNSF through the project CRSII2\_147609 and of the Spanish Ministerio de Ciencia e Innovaci\'on (MICINN) through Grant IJCI-2017-34300. F.Gh. acknowledges the partial support by Deutsche Forschungsgemeinschaft (DFG) under the grant project: 345463468 (idonate). Graphics have been prepared using the \textsc{matplotlib python package}~\cite{hunter-matplotlib-2007}.


\bibliography{main}

\end{document}